\documentclass[aps,pra,preprint,showpacs]{revtex4}
\usepackage{amsmath}
\numberwithin{equation}{section}% of amsmath
\usepackage{graphicx}
\usepackage{rotating}% allows 90°-rotated tables
\usepackage{comment}% To comment out long sections
\usepackage{ifpdf} % ensures proper latex and pdflatex output
\newcommand{\Zs}{Z^{\displaystyle *}} % symbol Z-star
\newcommand{\ssp}{\extracolsep{5pt}}% column separation in tables
\newcommand{\bsp}{\extracolsep{15pt}}
\newcommand{\zsp}{\extracolsep{0pt}}
\newcommand{\m}{\hphantom{$-$}} % aligns negative and positive numbers

\begin{document}
\title{
On various validity criteria for the configuration average 
in collisional-radiative codes
}% Original title was: "About various..." - postprint
\author{M. Poirier}
\affiliation{
Commissariat \`a l'\'Energie Atomique, 
Service ``Photons, Atomes et Mol\'ecules'', 
Centre d'\'Etudes de Saclay, 
F91191 Gif-sur-Yvette \textsc{Cedex} France }
\date{\today}
\begin{abstract}
The characterization of out-of-local-thermal-equilibrium plasmas 
requires the use of collisional-radiative kinetic equations. 
This leads to the solution of large linear systems, for which 
statistical treatments such as configuration average may bring 
considerable simplification. 
In order to check the validity of this procedure, a criterion based 
on the comparison between a partial-rate systems and the Saha-%
Boltzmann solution is discussed in detail here. Several forms of 
this criterion are discussed. The interest of these variants is 
that they involve each type of relevant transition (collisional or 
radiative), which allows one to check separately the influence of 
each of these processes on the configuration-average validity. 
The method is illustrated by a charge-distribution analysis in 
carbon and neon plasmas. Finally, it is demonstrated that when 
the energy dispersion of every populated configuration is 
smaller than the electron thermal energy, the proposed criterion 
is fulfilled in each of its forms. 
\end{abstract}
\maketitle

\section{Introduction}
Absorption and emission spectra in warm dense plasmas often exhibit 
broad structures described as unresolved transition arrays (UTA) 
\cite{Bau79,Bau82,Bau88}, spin-orbit split arrays (SOSA) 
\cite{Bau85} or supertransition arrays (STA) \cite{Bar89}. In 
order to account for huge numbers of transitions, one usually resort 
to statistical treatments based on configuration or 
superconfiguration average \cite{Bar97,Pey99,Bar00,Pey01,Bau04}. 
However, such an average may lack the accuracy required to describe 
isolated levels or lines. A possible improvement lies in the 
definition of effective temperatures inside each subset 
(configuration or superconfiguration), the detailed level 
populations inside a given subset being derived from the population 
of this subset and from these temperatures. The determination of 
such temperatures has been discussed in a series of papers 
\cite{Bus93,Bau00,Bau04,Han06}. Alternatively, one may consider 
hybrid models involving both fine-structure levels and 
configuration-averaged levels \cite{Han07}. 

In addition to the complexity of the spectra, in many situations
such as those prevailing in laser- or discharge-produced plasmas 
devised from extreme-UV production \cite{Nis06,OSu06,ARa06}, 
the electron density is not large enough to ensure local thermal 
equilibrium (LTE) through electron-ion collisions. One must then 
use collisional-radiative models: a set of pertinent transition 
rates is computed and stationary or time-dependant rate equations 
are then solved. A series of such models --- among which ATOMIC  
\cite{Hak06}, ATOM3R-OP \cite{Rod06}, AVERROES \cite{Pey01}, 
FLYCHK \cite{Chu05}, MOST \cite{Bau04}, SCAALP \cite{Fau01}, 
SCROLL \cite{Bar97,Bar00} --- have been developed and  
benchmarked in the NLTE workshops \cite{Bow03,Bow06}. 

While allowing for an accurate description of plasmas in a wide 
range of conditions, the collisional-radiative codes suffer from a 
practical intractability when large numbers of levels --- thousands 
or more --- are involved. On must then resort to the above 
mentioned average methods. However, this averaging procedure must 
be validated. In a preceding paper \cite{Poi07}, we proposed a 
criterion based on the solution of a partial-rate system and showed 
its efficiency in checking the validity of the configuration 
average (CA) procedure. This partial-rate system involved 
electron-impact excitation and ionization, plus the reverse 
processes. Nevertheless, since this criterion only involves 
collisional rates, it is desirable to implement a procedure that 
includes all types of transition processes. To this respect, a 
generalization of this method is proposed here, where every 
process is included in the validity check. This allows us to 
diagnose inhomogeneities in all kind of rates. 

In section \ref{sec:CAvalchk}, the general procedure for checking 
CA validity is presented. Section \ref{sec:radrate} deals with the 
case where the partial rate system contains radiative processes 
only. The CA-diagnostic method in several of its variants is then 
illustrated by detailed-versus-CA computations in carbon and neon 
plasmas (section \ref{sec:tstCNe}). Properties of the averaged rate 
system in two particular cases are analyzed in appendices 
\ref{app:prprat} and \ref{app:mrconflT}.

\section{A validity check based on detailed balance for the 
configuration-average procedure}\label{sec:CAvalchk}

\subsection{Detailed rate equations}
Let us consider an homogeneous plasma where electrons are assumed to 
be at thermal equilibrium with temperature $T_\text{e}$. The ions, 
with less collisional interactions are out of thermodynamic 
equilibrium when the electron density $N_\text{e}$ is low. No outer 
electromagnetic field is considered, which amounts to deal with 
optically thin media or in zero-temperature radiation fields. 
Ion-ion collisions and electron free-free transitions are not 
included, as they weakly affect the population transfers 
considered here. 

The included processes are radiative deexcitation, collisional 
ionization, three-body recombination, collisional excitation and 
deexcitation, radiative recombination, autoionization and 
dielectronic recombination. In the absence of any external field, 
the radiative deexcitation is unbalanced by the reverse 
process (photoabsorption), and radiative recombination is not 
balanced by photoionization. Accordingly, stimulated emission 
(bound-bound) and stimulated radiative recombination (free-bound) 
are not included either. 

In stationary regimes, plasma properties such as the charge 
distribution, the average internal energy, or the radiative 
parameters (opacity, emissivity, radiative losses) follow from the 
solution of the so-called collisional-radiative detailed rate 
equations 
\begin{equation}
	\frac{dN_i}{dt} = -\sum_{\substack{j\\j\ne i}} R_{ij} N_i 
	+\sum_{\substack{j\\j\ne i}} R_{ji} N_j = 0	\label{eqn:detrate}
\end{equation}
where $R_{ij}$ stands for the sum of all transition rates from $i$ 
to $j$.

\subsection{Configuration-average procedure}
Assuming the ionic level $i$ (resp. $j$) belong to configuration 
$\alpha$ (resp. $\beta$), the \textit{average rate} from $\alpha$ 
to $\beta$ is defined as 
\begin{equation}
	R_{\alpha\beta} = \frac1{g_\alpha} \sum_{\substack{i\in\alpha\\
	j\in\beta}} g_iR_{ij}, \text{\quad with\ }
	g_\alpha = \sum_{i\in\alpha}g_i\label{eqn:avratdef}
\end{equation}
where $g_i$ is the $i$-level degeneracy, and $g_\alpha$ is the 
configuration degeneracy. If the $N_i$ are solutions of the 
detailed system (\ref{eqn:detrate}), the total populations 
$N_\alpha = \sum_{i\in\alpha}N_i$ are usually not solution of the 
stationary configuration-averaged rate equations
\begin{equation}
	\frac{dN_\alpha}{dt} = -\sum_{\substack{\beta\\\beta\ne\alpha}} 
	R_{\alpha\beta} N_\alpha +\sum_{\substack{\beta\\\beta\ne\alpha}} 
	R_{\beta\alpha} N_\beta = 0	\label{eqn:avrateq}
\end{equation}
because the equations (\ref{eqn:detrate}) involve the product $RN$ 
which is nonlinear. However (see Appendix \ref{app:prprat}) one can 
derive that the sum of the detailed level populations is indeed a 
solution of the CA equation (\ref{eqn:avrateq}) in the special case 
where the transition rates $R_{ij}$ inside a pair of given 
configurations is simply proportional to the final degeneracy $g_j$.

\subsection{Validity criteria for the configuration average derived 
from detailed balance} \label{ssc:valCAdb}
Intuitively the CA procedure validity should require that the 
average energy dispersion (rms) must be less than the thermal 
electron energy
\begin{equation}
	\sum_\alpha N_\alpha \Delta E_\alpha \ll T_\text{e} 
	\label{eqn:avEdispcr}
\end{equation}
where $\Delta E_\alpha$ is the rms energy dispersion of the levels 
belonging to the configuration $\alpha$, its population $N_\alpha$ 
being normalized according to $\sum_\alpha N_\alpha=1$. 

However it has been established in the carbon test case, principally 
for low $T_\text{e}$, that the (\ref{eqn:avEdispcr}) criterion may 
severely fail \cite{Poi07}. Therefore a different criterion must be 
elaborated. 
It stems from the \textit{detailed-balance} (or microreversibility) 
principle on transitions between any pair $i,j$ of (detailed) levels
\begin{equation}
	N_i^\text{SB} R_{ij}^{(p)} = N_j^\text{SB} R_{ji}^{(q)}
	\label{eqn:detbal}
\end{equation}
$R_{ij}^{(p)}$ being the rate of any transition from $i$ to $j$ 
and $R_{ji}^{(q)}$ the rate for the inverse process. In 
equation~(\ref{eqn:detbal}), $N_i^\text{SB}$ is the Saha-Boltzmann 
(i.e., local thermal equilibrium) population of the $i$ level, 
obeying 
\begin{equation}
	N_j^\text{SB}/N_i^\text{SB} = 
	 \Theta^s \frac{g_j}{g_i} \exp(-E_{ji}/T_\text{e})\label{eqn:SBdet}
\end{equation}
where $E_{ji}=E_j-E_i$ is the transition energy between $i$ (net 
charge $\Zs$) and $j$ (net charge $\Zs\!\!+\!\!s$), 
$\Theta=2/(N_\text{e}\lambda_\text{th}^3)$, $N_\text{e}$ being the 
electron density and $\lambda_\text{th}$ the thermal wavelength 
$h/(2\pi m_\text{e} T_\text{e})^{1/2}$.

In the case where the energy dispersion in \emph{every} 
configuration is much smaller than $T_\text{e}$, 
\begin{equation}
	\Delta E_\alpha \ll T_\text{e} \text{\quad for all }\alpha
	\label{eqn:mxEdispcr}
\end{equation}
it can be demonstrated (see Appendix \ref{app:mrconflT}) that 
microreversibility holds for configurations too, i.e., that one 
has $N_\alpha^\text{SB} R_{\alpha\beta}^{(p)} = N_\beta^\text{SB} 
R_{\beta\alpha}^{(q)}$ for every pair of configurations. It 
implies that the configuration population distribution arising 
from the solution of the averaged microreversible rate equations 
agrees with Saha-Boltzmann law.

The nature of the processes $(p)$ and $(q)$ involved in 
equation (\ref{eqn:detbal}) has not been discussed yet. A previous 
work \cite{Poi07} had restricted the discussion to the case where 
$R_{ij}$ are the \emph{collisional} rates only. Here we want to 
compare the microreversibility criteria derived when these 
processes are
\begin{itemize}
	\item collisional excitation and ionization plus inverse processes, 
	as in \cite{Poi07}, hereafter named ``collisional'' case;
	\item photoexcitation and photoionization in a fictive Planckian 
	field, as discussed in the next section (``radiative'' case);
	\item both above processes plus autoionization and dielectronic 
	recombination (``complete'' case).
\end{itemize}
Let us notice that this last case is not identical to the usual 
collisional-radiative case, since it involves a fictive Planckian 
field. 

The interest of these various criteria is that they allow to check 
the dispersion inside a given pair of configurations of the 
probabilities for \emph{each} kind of process. This helps in 
determining which process is responsible for a breakdown of CA. 
For instance, as noted before \cite{Bus07}, collisional excitation 
cross sections may be difficult to derive from a simple fit 
formula such as those by Goett \textit{et al} \cite{Goe80}, while 
radiative transition rates are more regular. It must be noted that, 
in the CA case, the computational effort represented by these 
tests only amounts to the solution of one additional linear system 
for each criterion, since the transition rates have 
already been computed to solve the collisional-radiative problem. 
When one considers CA transitions, these matrix inversions 
represent little extra computation versus the evaluation of the 
various rates. To sum up, the proposed procedure consists in the 
following steps: computation of the detailed rates for each process 
with a suitable atomic code, followed by the averaging 
(\ref{eqn:avratdef}); resolution of the usual CA collisional-
radiative system (\ref{eqn:avrateq}); resolution of one of the 
\emph{modified} rate-equation systems as enumerated above and 
comparison with the Saha-Boltzmann solution. This comparison 
provides a direct indication of the dispersion of the transition 
probabilities within each pair of configurations, and thus a 
``figure of merit'' of the CA approximation.

The above-mentioned modified-rate systems may also be useful in the 
\emph{detailed} case as check for the matrix-inversion algorithm 
used: this is illustrated in subsection \ref{ssc:murevdet}.

\section{Rate equations involving radiative processes in a fictive 
Planckian field}\label{sec:radrate}

The purpose of this section is to derive a rate-equation system 
involving radiative transitions obeying to the microreversibility 
principle. This implies that the solution of this system must be 
Saha-Boltzmann in the detailed case (because Saha-Boltzmann 
solution does satisfy the rate equations and because the solution 
is assumed unique). In order to allow for detailed balance, one 
must consider here a \textit{fictive electromagnetic field with a 
Planckian distribution at a temperature} $T_\text{r}=T_\text{e}$.

\subsection{Definition of the bound-bound radiative rates}
\label{sec:bndbnd}
If the levels $i$ and $j$ correspond to the same charge state and 
$E_{ji}>0$, the absorption rate from $i$ to $j$ in a fictive 
field of spectral density $u_\nu$ is related to the spontaneous 
emission rate $A_{ji}$ according to the Einstein relations
\begin{equation}
	B_{ij}u_\nu = \frac{g_j}{g_i} A_{ji} \frac{h^2c^3}{8\pi E_{ji}^3} 
	u_\nu\\
	=\frac{g_j}{g_i} \frac{A_{ji}}{e^{E_{ji}/T_\text{e}}-1}
	\label{eqn:fictabs}
\end{equation}
where the density $u_\nu$ is evaluated at the transition frequency 
$E_{ji}/h$. In the same way, the \textit{stimulated emission rate} 
in the $u_\nu$ field is, using again Einstein relations,
\begin{equation}
	B_{ji}u_\nu = \frac{g_i}{g_j} B_{ij}u_\nu =
	\frac{A_{ji}}{e^{E_{ji}/T_\text{e}}-1}.\label{eqn:fictste}
\end{equation}
The relations (\ref{eqn:fictabs}, \ref{eqn:fictste}) allow one to 
derive the fictive absorption and stimulated emission rates 
(involved in the ``radiative-rate'' system) from the known 
spontaneous emission coefficients $A_{ji}$.

\subsection{Definition of the bound-free radiative rates}
\label{sec:bndfre}
Assuming that $i$ is a level of an ion with net charge $\Zs$, 
$j$ with net charge $\Zs\!+\!1$, photoionization and 
recombination processes for photon energies between $h\nu$ and 
$h(\nu+\delta\nu)$ are described by the kinetic equation
\begin{equation}
	\delta\frac{dN_j}{dt}=\left(r_{ij}^\text{pi}u_\nu N_i
	 -r_{ji}^\text{rr}N_eN_j-r_{ji}^\text{srr}u_\nu N_eN_j\right)
	 \delta\nu\label{eqn:pirrrrs}
\end{equation}
where $r^\text{pi}$, $r^\text{rr}$, $r^\text{srr}$ are the 
photoionization, radiative recombination and stimulated radiative 
recombination coefficients, which are (supposedly known) functions 
of the incident wavelength. 

When one solves a \textit{zero-field} collisional-radiative system, 
one integrates the radiative recombination coefficient 
$r_{ji}^\text{rr}$ over the photon energy and get the rate noted as 
$R_{ji}^\text{rr}$. To deal with a situation of a Planckian field 
in equilibrium with the plasma, it would be necessary to perform 
the integration of the photoionization and stimulated recombination 
coefficients in (\ref{eqn:pirrrrs}) on the $u_\nu$ Planck 
distribution, or on the electron Maxwellian distribution, the 
electron energy $\varepsilon$ being related to the field frequency 
through $E_i+h\nu=E_j+\varepsilon$. However this integration may 
be avoided since one is interested here not in the determination 
of real photoionization or stimulated recombination rates but in a 
system involving bound-bound and bound-free radiative rates, the 
solution of which is the Saha-Boltzmann distribution. One therefore 
introduces a unique spectral energy $u_\nu$ computed \emph{at 
threshold} $\nu=E_{ji}/h$ while in general one has $\nu\ge 
E_{ji}/h$. Then, in stationary regime, one changes equation 
(\ref{eqn:pirrrrs}) into the solution \textit{at threshold}
\begin{equation}
\begin{split}
	u_\nu(E_{ji}/h) &= \frac{R_{ji}^\text{rr}N_eN_j} 
	      {R_{ij}^\text{pi}N_i - R_{ji}^\text{srr}N_eN_j}\\
	&= \frac{R_{ji}^\text{rr}}{(g_i/g_j)R_{ij}^\text{pi}
	   e^{E_{ji}/T_\text{e}} \lambda_\text{th}^3/2 - R_{ji}^\text{srr}}
\end{split}
\end{equation}
if the Saha-Boltzmann relation (\ref{eqn:SBdet}) is explicitly used. 
From the already known recombination rate $R_{ji}^\text{rr}$, the 
equations 
\begin{align}
 R_{ij}^\text{pi}u_\nu(E_{ji}/h)&= \frac2{\lambda_\text{th}^3}
 \frac{g_j}{g_i}\frac{R_{ji}^\text{rr}}{e^{E_{ji}/T_\text{e}}-1}\\ R_{ji}^\text{srr}u_\nu(E_{ji}/h)&=
 \frac{R_{ji}^\text{rr}}{\displaystyle e^{E_{ji}/T_\text{e}}-1}
\end{align}
provide ``fictive'' photoionization and stimulated recombination 
rates respectively. The above defined $R_{ij}^\text{pi}$ and 
$R_{ji}^\text{srr}$ are not necessarily accurate values of the 
real photoionization and stimulated recombination coefficients, 
but they have the two required properties:
\begin{itemize}
	\item they depend on the known radiative recombination rate 
	$R_{ji}^\text{rr}$ in a simple way
	\item the solution of the detailed modified rate equations in this 
	fictive Planckian field including bound-bound and bound-free 
	rates as defined in subsections \ref{sec:bndbnd} and 
	\ref{sec:bndfre} is by construction the Saha-Boltzmann solution.
\end{itemize}
In the rather usual case where the photoionization cross-sections 
rapidly decrease at threshold, the above values are also realistic 
approximations of the photoionization and stimulated recombination 
rates in a Planckian field.

\section{A comparison of various partial rate equation systems}
\label{sec:tstCNe}
\subsection{Description of the calculations}
In order to illustrate the proposals of the previous sections, a 
detailed and configuration-average collisional-radiative analysis 
has been performed in carbon and neon plasmas, including all 
possible charge states in order to correctly describe a large 
range of temperatures and electron densities. The atomic and 
collisional computations have been done using the HULLAC (Hebrew 
University Lawrence Livermore Code) suite \cite{Bar01}. This code 
includes full account of configuration interaction, of particular 
importance, e.g., in plasmas devised for extreme-UV generation 
\cite{Gil03,OSu06,Nis06}. It allows for a fast computation of 
collisional cross-sections using the factorization-interpolation 
method.

The included configurations in C are the same as in our previous 
work \cite{Poi07}. The configurations in Ne, chosen according 
to the same criterion, are enumerated in Table~\ref{tab:confNe}.
This gives rise to 150 configurations and 1782 levels 
in C, where the computation time remain reasonable while allowing 
for a valid check of the present criteria for the configuration 
average procedure. In Ne, the 4638-level case give rise to 
longer calculations which provide a serious accuracy check for 
the proposed criterion. One may estimate that the maximum value 
of the principal quantum number $n=5$ may be too low to describe 
certain quantities, e.g., radiative losses, which depend mainly 
on excited level populations. Several processes such as 
dielectronic recombination may have a large cross-section for 
high $n$. However the present work is mainly aimed at checking 
the operation of validity criteria --- which of course would 
apply to higher $n$ --- and not at providing a reference NLTE 
computation. Furthermore, a test in carbon for $N_\text{e}=
10^{16}\text{ cm}^{-3}$ has established that increasing the 
maximum $n$ from 5 to 6 changes the average ionization from 
$\left<\Zs\right>=3.927$ to 3.937 if $T_\text{e}=10\text{ eV}$ 
and from 0.934 to 0.935 if $T_\text{e}=1\text{ eV}$, and the 
computed radiative losses exhibit a less-than-15\% change. 
This indicates an acceptable accuracy while keeping the amount 
of computations --- which can be considerable when dealing with 
detailed-level check --- moderate.

The neon computation involves 541~713 radiative deexcitation 
rates, 1~532~357 collisional excitation rates, 2~431~784 
collisional ionization rates, 1~144~559 radiative recombination 
rates, and 23~321 autoionization rates. As discussed previously 
\cite{Poi07}, some transition rates are poorly determined in 
HULLAC. For instance few collisional excitation rates are sensitive 
to inaccuracy in the Sampson fit used \cite{Goe80,Bus07}. It turns 
out that, at 30 eV, 6~433 collisional excitation rates (0.4\%) are 
negative, 3~402 (0.3\%) collisional ionization rates are singular 
\cite{Poi07} and 24 (0.1\%) autoionization rates are abnormally 
large. Since these fractions remain small, such irregular values 
may be simply cancelled. % canceled -> cancelled - postprint

In order to estimate the influence of such cancellation of some % cancellation postprint 
collisional excitation rates --- for which the higher percentage 
of unexpected values is observed --- a simple check has been 
performed. It consists in substituting to these unknown rates 
simple analytical expressions relating the collisional deexcitation 
cross-section to the radiative deexcitation rate \cite{vRe62,Mew72}. 
For instance, in carbon at 10 eV, 8~190 collisional excitation rates 
out of 465~083 (1.8\%) are negative when computed with HULLAC. Using 
a Van Regemorter-type formula with an average Gaunt factor equal to 
0.2, one may estimate 4~583 additional rates (56\% of the negative 
values); the other rates correspond to transitions forbidden by 
electric-dipole selection rules, for which a null cross-section 
seems to be acceptable. Doing this, the average ionization degree at 
$N_\text{e}=10^{16}\text{ cm}^{-3}$ is $\left<\Zs\right>=3.9275$ 
while it was 3.9273 when cancelling all negative excitation rates.  
Therefore, as a rule, the cancellation of few irregular rates bears  
very little consequence. % cancelling, cancellation - postprint

A comparison with neon detailed and superconfiguration-averaged 
values published by Hansen \textit{et al} \cite{Han06} is presented 
in Table~\ref{tab:compSH}. The DLA values also results from a 
HULLAC-based detailed calculation, while MOST is a 
superconfiguration code based on a self-consistent field 
\cite{Pey01,Bau04}. The ``hybrid'' model combines the power of 
statistical average with the accuracy of detailed models and is 
based on the Flexible Atomic Code \cite{Gu03}; contrary to other 
computations, it includes the continuum lowering \cite{Han07}. 
The agreement between these values and the present one is 
satisfactory. As discussed by Hansen \textit{et al}, the differences 
arise from the diversity in the underlying atomic data rather than 
from the averaging process. The more significant departure with the 
hybrid model at high electron densities originates certainly in the 
account for continuum lowering.

\subsection{Using the microreversibility principle in the detailed 
case}\label{ssc:murevdet}
The solution of the detailed system (\ref{eqn:detrate}) with 
modified  (microreversible) rates as discussed in subsection 
\ref{ssc:valCAdb} provides a useful accuracy check when 
comparison is made with Saha-Boltzmann.

This is illustrated for a 10-eV neon plasma in Table 
\ref{tab:acthsbNe}. The average ionization degree is computed for 
the various rate systems detailed at the end of subsection 
\ref{ssc:valCAdb}. One observes that the ``collisional'', 
``radiative'', and ``complete'' rate equations provide a 
$\left<\Zs\right>$-value in close agreement with Saha-Boltzmann 
equation, usually agreeing within the nine digits, with a small 
degradation for large electron densities. 
It is essential to note that this $N\times N$ ($N=4637$) matrix 
inversion is performed with a remarkable accuracy, the Gauss 
elimination used resulting in a number of floating-point operations 
scaling as $N^3$. In addition to the $\left<\Zs\right>$ comparison, 
the tests on the maximum population difference (e.g., 
$\max|N_i\text{(coll)} - N_i\text{(SB)}|$ for $1\le i\le4638$) 
displayed in this table confirms the accuracy of the present 
rate-equation solution.

\subsection{Configuration average in the high-temperature regime}
Let us first consider high-$T_\text{e}$ situations where CA is 
supposed to be valid. 
The CA-validity test in its different forms has been first applied 
to a carbon plasma at 10 eV. As seen in Table~\ref{tab:acthsbC}, 
it turns out that CA is then as a rule acceptable, except for 
$N_\text{e}>10^{21}\text{ cm}^{-3}$ where up to 50\% divergence 
is observed. In this case, two of the tests proposed here reveal 
the CA-validity breakdown. While for $N_e=10^{22}$ the average 
dispersion $\Delta E_\alpha$ is 2 eV, well below the thermal 
energy, the $\left<\Zs\right>$ value computed with ``collisional'' 
rates (1.053) or ``complete'' rates (1.070) differ significantly 
from the Saha-Boltzmann ionization (0.718). In this case the ratios 
$\left<\Zs_\text{coll}\right>/\left<\Zs_\text{SB}\right>=1.47$ or 
$\left<\Zs_\text{all}\right>/\left<\Zs_\text{SB}\right>=1.49$ are 
%--- though accidentally --- 
acceptable approximations for the CA/detailed ratio 
$\left<\Zs_\text{CA}\right>/\left<\Zs_\text{det}\right>=1.50$. 
However, the ``radiative'' test value 
$\left<\Zs_\text{rad}\right>$ remain much closer to Saha-Boltzmann, 
the maximum discrepancy being 
$\left<\Zs_\text{rad}\right>/\left<\Zs_\text{SB}\right> = 0.96$ at 
$N_\text{e}=10^{22}\text{ cm}^{-3}$. Here the ``radiative'' 
test is less sensitive to irregularities in autoionization and 
collisional ionization previously noticed \cite{Poi07} and 
therefore it does not detect accurately the CA-validity breakdown. 
However a safe way to control the CA validity is to compare the 
charge $\left<\Zs_\text{all}\right>$ (and not 
$\left<\Zs_\text{coll}\right>$) to $\left<\Zs_\text{SB}\right>$ 
because the latter quantity is sensitive to irregularities in all 
kinds of processes.

A comparison between detailed and CA results in a 30 eV-neon plasma 
is presented in Table~\ref{tab:CAchkNeT30Z}. For this temperature 
the CA approximation is acceptable in the whole density range 
investigated, the error varying from 3\% to 1\%. 
However, the energy criterion (\ref{eqn:avEdispcr}), though 
globally satisfied, does not give an accurate picture of the CA 
validity. For increasing density numbers, the CA approximation 
improves while the criterion (\ref{eqn:avEdispcr}) 
is less verified. While in this table the energy dispersion 
$\Delta E_\alpha$ is evaluated using configuration populations 
$N_\alpha$ derived from the CA-collisional-radiative solution, 
using $N_\alpha$ from Saha-Boltzmann equation would increase 
monotonically from $10^{-10}$ eV to 5.7 eV in this density range. 

Nevertheless in the 30-eV neon case, the ratio 
$\left<\Zs_\text{all}\right>/\left<\Zs_\text{SB}\right>$ cannot 
provide a uniform test for CA validity. Noticeably at this 
temperature, for low $N_\text{e}$ the most probable ion state at 
thermal equilibrium (very different from the collisional-radiative 
solution) is the closed-shell ($1s^2$) Ne\textsc{ix} ground state. 
As a consequence, the ratio $\left<\Zs_\text{all}\right>/
\left<\Zs_\text{SB}\right>$ is very close to unity while the CA 
and detailed charge differ by about 3\%. In order to monitor this 
difference, a complementary global information is given by the 
various central moments of the population distribution
\begin{equation}
	m_t = \left<(\Zs-\left<\Zs\right>)^t\right> = 
	\sum_{0\le k\le Z} P_k (k-\left<\Zs\right>)^t\label{eqn:def_mt}
\end{equation}
where $P_k$ is the ion population of net charge $k$, and $t=2 %space 
\text{ or }3$ here. The corresponding results for a 30-eV neon plasma 
are presented in Tables \ref{tab:CAchkNeT30m2} and 
\ref{tab:CAchkNeT30m3}. 
One may check that, at $N_\text{e}=10^{12}\text{ cm}^{-3}$ 
for instance the ratio $m_2\text{(coll)}/m_2\text{(SB)}$ is 0.93 
(the ratio formed with ``radiative'' or ``complete'' $m_2$ being 
very similar) while the $m_2\text{(CA)}/m_2\text{(det)}$ is 
0.86. Even (much) stronger discrepancies could be obtained 
using third central moments. Therefore the use of second (and 
possibly third) central moments together with the average charge 
and comparison of ``collisional'' or ``complete'' data to 
Saha-Boltzmann data provide as a rule an efficient and uniform 
test for the CA validity.

\subsection{Configuration average in the low-temperature case}
\begin{figure}[htbp]
\caption{Average ionization $\left<\Zs\right>$ and second central 
moment $m_2 = \left<(\Zs-\left<\Zs\right>)^2\right>$ for a 10-eV 
neon plasma, in the detailed level scheme (symbols) and in 
configuration average (lines). Populations are computed either 
with the collisional-radiative code or with the Saha-Boltzmann 
equation.\label{fig:Zm_m2_Ne_T10}}
	\centering
		\includegraphics[scale=0.70,angle=0]{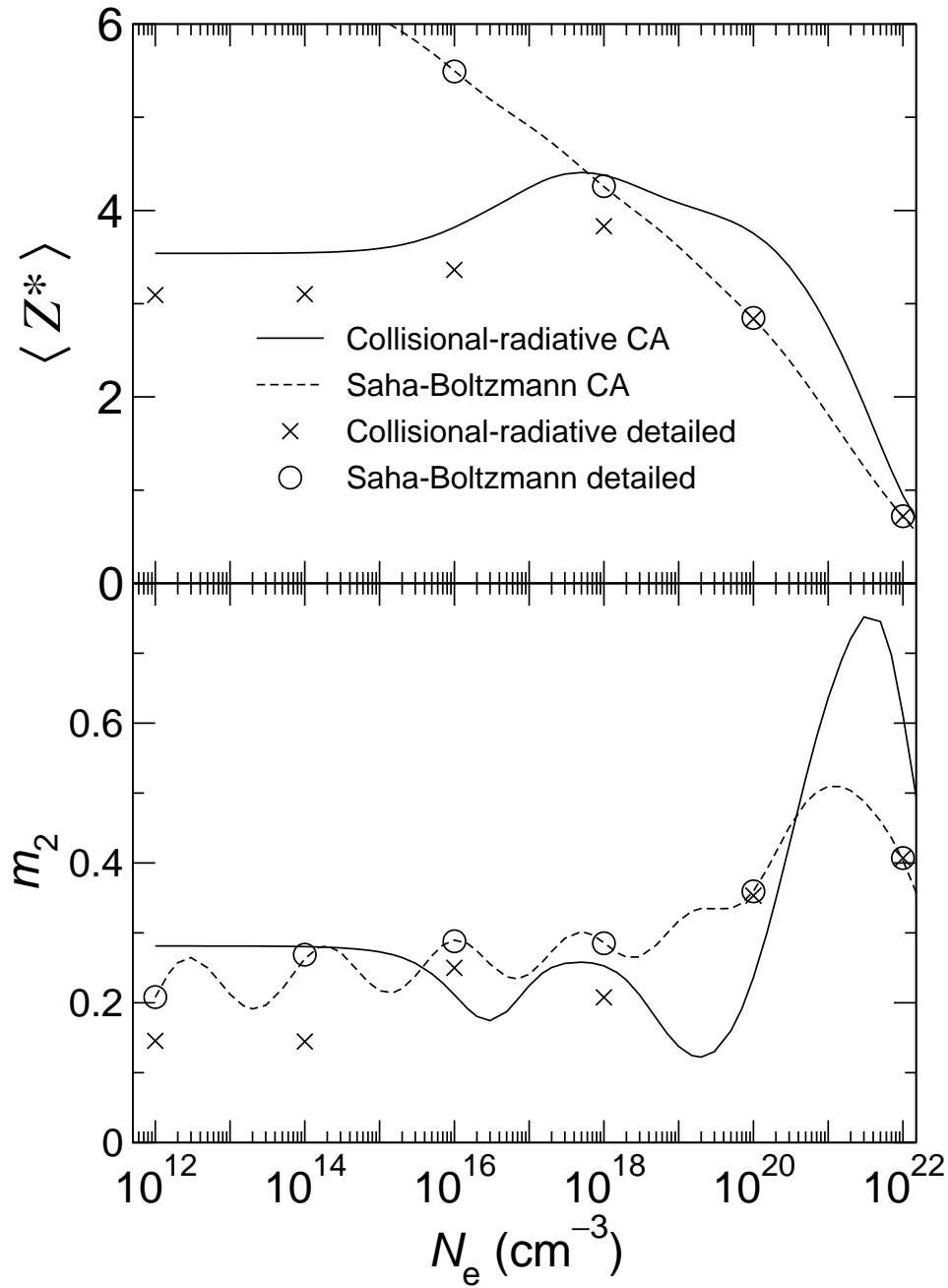}
\end{figure}

In figure \ref{fig:Zm_m2_Ne_T10} are presented the average charge 
$\left<\Zs\right>$ and the second central moment $m_2$ 
(\ref{eqn:def_mt}) for a 10-eV neon plasma, both in the 
detailed-level scheme and in the CA approximation. The figure 
includes Saha-Boltzmann results too. For  densities large enough 
(at least $10^{20}\text{ electrons/cm}^{3}$ ), the \emph{detailed} %cm^{-3} ->cm^3 postprint
data obtained with the collisional-radiative code (crosses) are 
identical to the Saha-Boltzmann values (circles), indicating that 
LTE is then reached. One notes that Saha-Boltzmann data are % notices -> notes - postprint
almost identical in the detailed (circles) and the CA case 
(dashed line). This indicates that the averaging on energies then 
performed in the Saha-Boltzmann equation bears very little 
influence on populations. Conversely, the CA (crosses) and detailed 
(solid line) results obtained with the collisional-radiative system 
are significantly different, indicating that the averaging 
procedure on rate equations is not of little consequence. 
As expected the CA validity worsens at low $T_\text{e}$, the 
difference being 
$\left<\Zs_\text{CA}\right> - \left<\Zs_\text{det}\right> = 0.92$ at 
$N_\text{e}=10^{20}\text{ cm}^{-3}$. As seen in the lower part of 
this figure, the second moment $m_2$ exhibits similar properties, 
e.g., concerning the LTE convergence in the detailed case and the 
collisional-radiative vs Saha-Boltzmann differences. One can easily 
check that, if few charge states are populated, when 
$\left<\Zs\right>$ is integer (resp. half-integer), $m_2$ is close 
to a minimum (resp. maximum). This explains the oscillations 
observed on this figure.

\begin{figure}[htbp]
\caption{Average configuration energy dispersion $\left<\Delta 
E\right> = \sum_\alpha N_\alpha \Delta E_\alpha$ in a $T_\text{e} 
=10\text{ eV}$ neon plasma. $\Delta E_\alpha$ is the energy 
dispersion inside the configuration $\alpha$, and $N_\alpha$ is the 
$\alpha$-population calculated with various rate-equation systems 
and with Saha-Boltzmann equation. The smaller these dispersions are 
versus $T_\text{e}$, the better is assumed to be the configuration 
average approximation.\label{fig:DeltaE_Ne_T10}}
	\centering
\ifpdf% pdflatex is true
	\includegraphics[scale=0.55, angle=0]{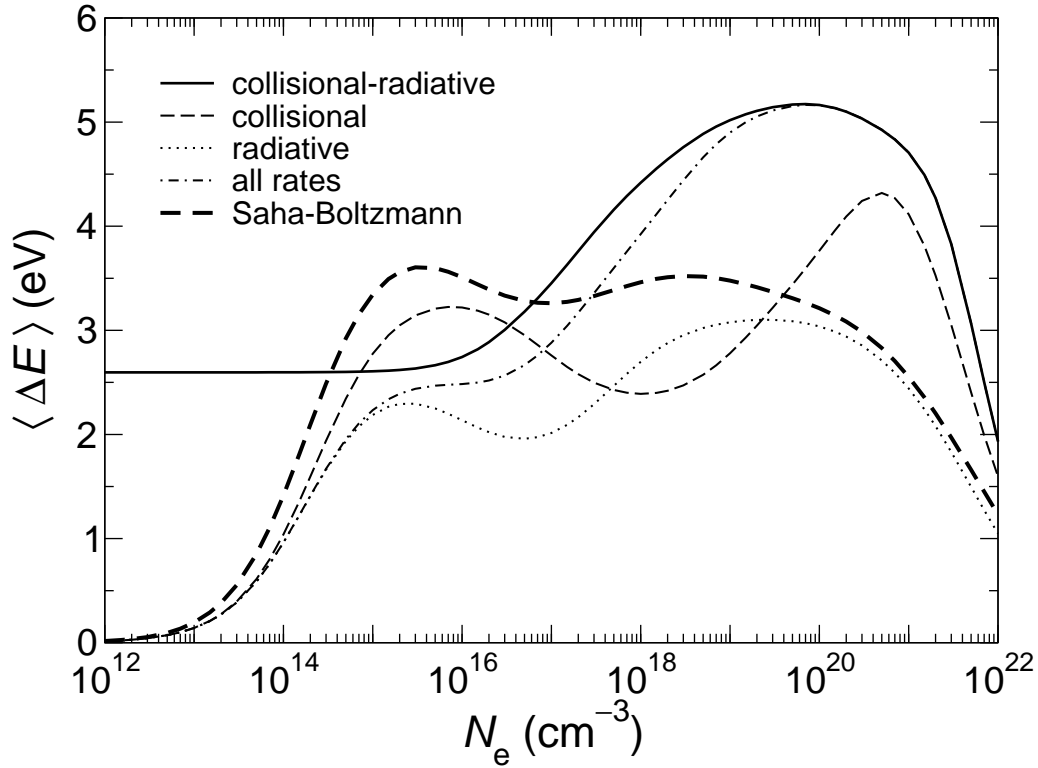}
\else% pdflatex is false
	\includegraphics[scale=0.55, angle=-90]{colradconf_DE_T10}
\fi

\end{figure}

In order to check the breakdown of CA validity \emph{without 
performing any detailed-level computation}, a first criterion 
is provided by the average-energy dispersion (\ref{eqn:avEdispcr}). 
The first member of this equation is plotted as a function of 
$N_\text{e}$ in figure \ref{fig:DeltaE_Ne_T10}. It appears that the 
energy dispersion $\left<\Delta E\right>$ is here rather large, in 
the 2--5 eV range, to be compared to $T_\text{e}=10\text{ eV}$. 
Furthermore, the dispersion calculated with the collisional-%
radiative system solution is always greater than 1.9 eV, and peaks 
at 5.2 eV for $N_\text{e}=10^{20}\text{ cm}^{-3}$ which is 
precisely the maximum of the CA versus detailed difference on 
$\left<\Zs\right>$. If one estimates that the criterion 
(\ref{eqn:avEdispcr}) is not fulfilled for $\left<\Delta 
E\right>=2.6\text{ eV}$, it has the expected behavior. The energy 
dispersion $\left<\Delta E\right>$ may also be computed 
with populations from the Saha-Boltzmann equation or from the 
modified rate equations proposed in subsection \ref{ssc:valCAdb}. 
As seen in figure \ref{fig:DeltaE_Ne_T10}, it is smaller than the 
dispersion from collisional-radiative solution (except in the 
$N_\text{e}\simeq10^{15}\text{--}10^{16}\text{ cm}^{-3}$ range) 
and particularly for electron densities well below $10^{13}\text{ 
cm}^{-3}$ where this quantity tends to zero. This can be explained 
as follows, e.g., for the ``collisional'' rate equations. 
Ignoring the collisional excitation and de-excitation processes 
that do not change the ionization degree, the ionization 
balance depends on the collisional ionization rate from 
the charge state $z$ to $z+1$, $R^\text{(ci)}N_z N_\text{e}$, 
and on the three-body recombination process from $z\!+\!1$ to $z$, 
$R^\text{(3br)}N_{z+1} N_\text{e}^2$. At equilibrium, this gives a 
population ratio $N_{z}/N_{z+1} \sim N_\text{e}R^\text{(3br)}/ 
R^\text{(ci)}$, vanishing at low $N_\text{e}$. 
Therefore the highest charges are then favored: it 
should be 10 at very low $N_\text{e}$, but since the closed-shell 
$1s^2$ Ne \textsc{ix} is very stable, one gets $\left<\Zs\right> = 
8$ over a large range of densities. A similar analysis may 
be performed for the ``radiative'' equilibrium where photoionization 
and radiative recombination are included. In these cases, since the 
mainly populated configuration is $1s^2$ with $\Delta E_\alpha=0$, 
the resulting average energy dispersion is very small. In the 
contrary, in the \emph{collisional-radiative case}, the low-density 
balance (coronal plasmas) is governed by collisional ionization 
and radiative recombination and then one gets a constant nonzero 
$N_z/N_{z+1}$, and therefore $\left<\Zs\right>\neq 8$. For 
this rate system, the average dispersion $\left<\Delta E\right>$ 
is nonzero, as seen in figure \ref{fig:DeltaE_Ne_T10}.

\begin{figure}[htbp]
\caption{Difference between the average charge calculated with 
various rate-equation systems and the average charge obtained from 
the Saha-Boltzmann equation. Calculations are performed in a 10-eV 
neon plasma in the configuration average scheme. 
\label{fig:DZ_Ne_T10}}
	\centering
\ifpdf% pdflatex is true
	\includegraphics[scale=0.55, angle=0]{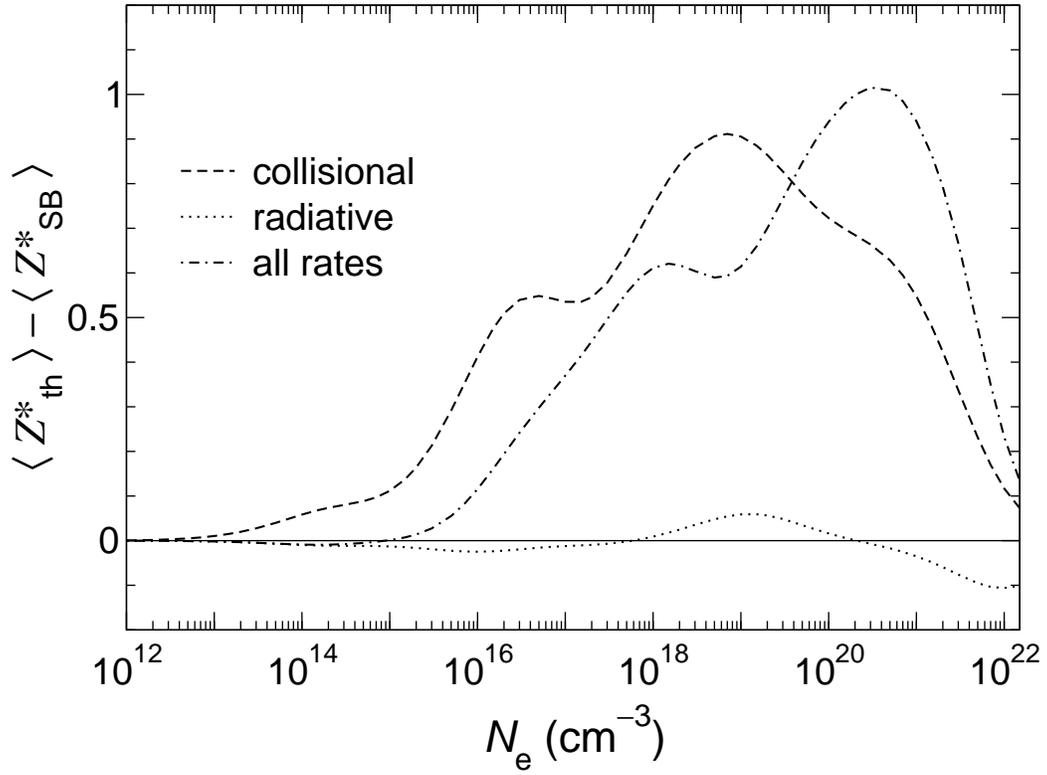}
\else% pdflatex is false
	\includegraphics[scale=0.55, angle=-90]{colradconf_DZ_T10}
\fi
\end{figure}

To check how the proposed criterion based on the solution of 
a partial rate equations behaves, one may consider the variation 
of $\left<\Zs_\text{th}\right>-\left<\Zs_\text{SB}\right>$ 
displayed in figure \ref{fig:DZ_Ne_T10}. It is expected that, when 
this difference is large, the CA approximation will fail. 
It first appears that the ``radiative'' solution as proposed in 
section \ref{sec:radrate} remain rather close to the 
Saha-Boltzmann solution, and cannot in this case provide a test 
for the CA validity. However, the ``complete'' solution involving 
all the microreversible rates (collisional, autoionization and 
radiative in a fictive Planck field) presents a maximum difference 
on $\left<\Zs_\text{all}\right>-\left<\Zs_\text{SB}\right>$ 
at $N_\text{e}=3\times10^{20}\text{ cm}^{-3}$, precisely where the 
CA-detailed departure is maximum. Furthermore, this 
$\left<\Zs_\text{all}\right>-\left<\Zs_\text{SB}\right>$ value, 
close to 1, provides then a reasonable quantitative estimate of the 
CA-detailed difference, at variance with the simple criterion 
(\ref{eqn:avEdispcr}). The drawback of the present criterion lies 
here once again the the low-density region, where all the computed 
$\left<\Zs\right>$ are close to 8, and therefore the differences 
plotted on figure \ref{fig:DZ_Ne_T10} tend to zero. 
In this particular case, one must resort to the usual energy 
criterion (\ref{eqn:avEdispcr}). This is an opposite situation to 
the carbon case \cite{Poi07}: for instance, if $N_\text{e}=
10^{12}\text{ cm}^{-3}$ and $T_\text{e}=1\text{ eV}$, while 
the energy criterion (\ref{eqn:avEdispcr}) is fulfilled by large, 
the average charge difference $\left<\Zs_\text{coll}\right>-
\left<\Zs_\text{SB}\right>$ is close to 1, indicating that CA 
approximation does fail.

An alternate analysis may be performed on the second or third 
central moments computed with the solution with rates as defined 
in section \ref{sec:radrate} and compared to the analogous 
moments from Saha-Boltzmann solution. It has been checked that 
the conclusions remain similar, with a maximum difference around 
$N_\text{e}=10^{21}\text{ cm}^{-3}$. Again the ``radiative'' 
solution remain closer to Saha-Boltzmann than the ``collisional'' 
or ``full'' solution.

\section{Conclusion}
It has been demonstrated that the configuration-average validity
for collisional-radiative rates, which leads to considerable 
simplification of these equation systems, may be controlled by 
comparison of several variants of a modified rate system with 
Saha-Boltzmann solution. These tests have been performed 
on the average ionization degree as well as the second and third 
central moments. The radiative test appears to be easier to verify 
than the collisional or complete tests: This demonstrates that 
the CA breakdown is induced mainly by large dispersions in 
collisional rates inside a given pair of configurations and not 
in radiative rates.
Tests have been performed in carbon and neon, where atomic and 
collisional data were provided by the HULLAC code. Going from C to 
Ne amounts to increase the number of levels from 1782 to 4638 
(and from 150 to 291 configurations) without impairing the 
efficiency of the proposed test. 
A detailed analysis in a carbon plasma has proven that failure of 
the CA at $T_\text{e}=1\text{ eV}$ may be correctly detected by 
the test based on the modified-rate system, while failure of CA 
in a 10 eV-neon plasma at low density is better diagnosed by the 
intuitive criterion based on the average energy dispersion 
inside a configuration. This situation prevails when the 
modified-rate equations solution is dominated by closed-shell 
ions such as Ne\textsc{ix}. Therefore, both criteria appear quite 
complementary. Nevertheless, the criterion based on the average 
charge difference may provide a semi-quantitative estimate of the 
CA-versus-detailed charge difference, at variance with the 
dispersion energy criterion. Further developments include the 
use of such systems in the determination of physically important 
quantities such as radiative losses, the opacity, and the 
emission of non-equilibrium plasmas \cite{Pey01,Chu06,Col06}. 
It is interesting to check how the various forms of the present 
criterion can control the relevance of CA approximation when 
dealing with such radiative properties. As a possible example 
of application, a configuration averaged code based on the HULLAC 
suite has been recently applied in our group to the EUV emission 
of xenon plasmas \cite{dGD06}.

\begin{acknowledgments}
The author gratefully acknowledges Dr. T. Blenski for constant 
support and Dr. S. Hansen for useful comments and for providing 
reference neon data. He is also indebted to the developers of the 
HULLAC code for making it available and to F. de Dortan for his 
assistance on the collisional-radiative codes. 
\end{acknowledgments}

\appendix
\section{Validity of the configuration average: an example
\label{app:prprat}}
In the special case where the rates inside every pair of 
configurations are proportional to the final level degeneracy
\begin{equation}
	R_{ij}=R_{\alpha\beta} \frac{g_j}{g_\beta} 
	\quad\forall i\in\alpha, j\in\beta,	\label{eqn:hyprate}
\end{equation}
the solution of the \textit{detailed-level} rate equations can be 
straightforwardly derived from the solution of the 
\textit{configuration-average} rate equations. Let us assume that 
the populations ${\bar{N_\alpha}}$ are the solutions of the CA 
rate equation (\ref{eqn:avrateq}). If we \textit{define} level 
populations according to
\begin{equation}
	\bar{N_i}=\bar{N_\alpha} \frac{g_i}{g_\alpha},\ 
	\bar{N_j}=\bar{N_\beta} \frac{g_j}{g_\beta} 
	\quad i\in\alpha, j\in\beta \label{eqn:poplpc}
\end{equation}
then it is easy to show that such populations obey the 
\textit{detailed} rate equation (\ref{eqn:detrate}). One has 
according to (\ref{eqn:hyprate}) and (\ref{eqn:poplpc})
\begin{equation}
%\begin{split}
	\sum_{\substack{j\\j\neq i}}\left(-R_{ij}\bar{N_i}+
	R_{ji}\bar{N_j}\right)=\sum_{\substack{j\\j\neq i}}\left( 
	 -R_{\alpha\beta}\frac{g_ig_j}{g_\alpha g_\beta}\bar{N_\alpha}
	 +R_{\beta\alpha}\frac{g_jg_i}{g_\beta g_\alpha}\bar{N_\beta}\right)
%\end{split}
\end{equation}
where $\alpha$ (resp. $\beta$) is the configuration containing $i$ 
(resp. $j$). In the above sum, when the level $j$ belongs to the 
same configuration as $i$ ($\beta=\alpha$), the $j$-term vanishes 
identically; when $j$ belongs to another configuration 
($\beta\neq\alpha$), the sum over $j\notin\alpha$ may be written 
$\sum_{\beta\neq\alpha}\sum_{j\in\beta}$ and then, since 
$\sum_{j\in\beta}g_j/g_\beta=1$, one gets after performing the 
$j$-sum
\begin{equation}
%\begin{split}
	\sum_{\substack{j\\j\neq i}}
	\left(-R_{ij}\bar{N_i}+R_{ji}\bar{N_j}\right)	=
	\frac{g_i}{g_\alpha} \sum_{\substack{\beta\\\beta\neq\alpha}}
	\left(-R_{\alpha\beta}\bar{N_\alpha}+R_{\beta\alpha}\bar{N_\beta}
	 \right)=0
%\end{split}
\end{equation}
because of the $\bar{N_\alpha}, \bar{N_\beta}$ satisfy the CA rate 
equations (\ref{eqn:avrateq}). In a case where (\ref{eqn:hyprate}) is 
fulfilled the microreversibility-based (or ``thermodynamic'') test 
checking the CA validity proposed in this paper is satisfied: the 
populations obtained with partial rate equations are identical to 
Saha-Boltzmann.

\section{Average of microreversible rates when the energy 
dispersion is small versus the electron thermal 
energy\label{app:mrconflT}}
In this appendix it is proved that, if the energy dispersion inside 
\textit{every} configuration is much smaller than the electron 
thermal energy (\ref{eqn:mxEdispcr}), then the microreversibility 
condition on detailed levels implies microreversibility on 
\textit{configuration-averaged} rates too. Let us assume that the 
processes $p$ and $q$ (e.g., collisional ionization and three-body 
recombination) obey the microreversibility condition
\begin{equation}
	R_{ij}^{(p)}N_i^\text{SB} = R_{ji}^{(q)}N_j^\text{SB}
\end{equation}
for every $i\in\alpha, j\in\beta$, where the populations 
$N_i^\text{SB}, N_j^\text{SB}$ obey the Saha-Boltzmann equation 
(\ref{eqn:SBdet}). Then, using the average rate definition 
(\ref{eqn:avratdef}) one gets
\newcommand{\tvi}{\vrule width 0pt height 2.25ex depth 1.0 ex}
\begin{equation}\begin{split}
 R_{\alpha\beta}^{(p)} &= \frac1{g_\alpha}
 \sum_{\substack{i\in\alpha\\j\in\beta}} g_i
% \frac{N_j^\text{SB}}{N_i^\text{SB}}
% \frac{N_j^\text{SB}}{\strut N_i^\text{SB}}
 \frac{N_j^\text{SB}}{\tvi N_i^\text{SB}}
 R_{ji}^{(q)}\\
 &=\frac{\Theta^s}{g_\alpha}\sum_{\substack{i\in\alpha\\
 j\in\beta}}g_j
 e^{(-E_{ji}/T_\text{e})}R_{ji}^{(q)} \label{eqn:calcar}
\end{split}\end{equation}
and \textit{if it is possible to identify the energy difference} 
$E_{ji}$ with the difference of the average configuration energy 
\begin{equation}
	\left| E_{ji} - E_{\beta\alpha} \right| \ll T_\text{e}
	\label{eqn:dispElT}
\end{equation}
then after substitution of $E_{\beta\alpha}$ to $E_{ji}$, the rate 
expression (\ref{eqn:calcar}) involves the Saha-Boltzmann ratio 
of the \textit{configuration populations}
\begin{equation}\begin{split}
 R_{\alpha\beta}^{(p)} &\simeq \frac{\Theta^s}{g_\alpha}
 e^{-E_{\beta\alpha}/T_\text{e}} \sum_{\substack{i\in\alpha\\
 j\in\beta}}g_j R_{ji}^{(q)}\\
 &= \frac{g_\beta}{g_\alpha}\Theta^s e^{-E_{\beta\alpha}/T_\text{e}}
 R_{\beta\alpha}^{(q)} = \frac{N_\beta^\text{SB}}{N_\alpha^\text{SB}}
 R_{\beta\alpha}^{(q)}\label{eqn:SBav}
\end{split}\end{equation}
where the Saha-Boltzmann populations $N_\alpha^\text{SB}$ computed 
with \textit{average energies} have been introduced. This means 
that, in this case, the microreversibility condition holds for 
\textit{configurations} too. However, conditions 
(\ref{eqn:mxEdispcr}) or  (\ref{eqn:dispElT}) assume that the 
maximum energy dispersion on \emph{every} pair of 
configurations must be smaller than $T_\text{e}$, which is a much 
stronger condition than (\ref{eqn:avEdispcr}) which only requires 
the \emph{average} energy dispersion to be much less than 
$T_\text{e}$.

\clearpage
%\bibliography{colrad2}% associated bib file
% Lists all entries in colrad2.bib
%\nocite{*}

%===========================%
%                           %
%  B I B L I O G R A P H Y  %
%  from the bbl file        %
%===========================%
%\newpage

\clearpage
%%%%%%%%%%%%%%%%%%%%%%%%%%%%%%%%%%%%%%%%%%%%%%%%%%%%%%%%%%%%%%%%%%
%%%%%%%%%%                    Tables                   %%%%%%%%%%%
%%%%%%%%%%%%%%%%%%%%%%%%%%%%%%%%%%%%%%%%%%%%%%%%%%%%%%%%%%%%%%%%%%
\section*{Tables}

\enlargethispage{1.5cm}
\begin{table}[htp]%[htbp]
\begin{center}
\caption{Number of configurations and levels considered for the 
collisional-radiative calculations in the neon plasma. For 
Ne\textsc{i} to Ne\textsc{viii}, the included configurations are 
$1s^2\{2s2p\}^k$, $1s^2\{2s2p\}^{k-1}Nl$, where $\{2s2p\}^j$ stands 
for $2s^22p^{j-2}, 2s2p^{j-1}, 2p^{j-2}$. The configurations 
considered in Ne\textsc{ix} are $1s^2\text{ and }1sNl$, those in 
Ne\textsc{x} are $Nl$. One has $N\le5, l\le N-1$ for each ion. 
The C\textsc{i}--C\textsc{vii} computations include configurations 
isoelectronic to Ne\textsc{v}--Ne\textsc{xi} respectively.
\label{tab:confNe}}
\bigskip
\begin{tabular}{r@{\quad}@{\extracolsep{5pt}} *{12}{c}}
\hline\hline
Ion & Ne\textsc{i} & Ne\textsc{ii} & Ne\textsc{iii} & Ne\textsc{iv} 
 & Ne\textsc{v} & Ne\textsc{vi} & Ne\textsc{vii} & Ne\textsc{viii} 
 & Ne\textsc{ix} & Ne\textsc{x} &  Ne\textsc{xi} & Total\\
\hline
Configurations & 25 & 38 & 39 & 39 & 39 & 39 & 27 & 14 & 15 & 15 & 1 & 291\\
Levels        & 157 &501 &994 &1204&1004&513 &166 & 24 & 49 & 25 & 1 &4638\\
\hline\hline
\end{tabular}
\end{center}
\end{table}
%\clearpage

\begin{table}[hbp]%[htbp]
\begin{center}
\caption{Average ionization level in a neon plasma 
at $T_e=25\text{ eV}$ and 50 eV. DLA is a detailed model, while MOST 
is based upon superconfiguration averaging \cite{Han06}. The code 
``hybrid'' is a partially detailed and partially averaged model 
based on FAC code \cite{Han07,Gu03}. The last 
two columns are the present results, in detailed and configuration 
average form.\label{tab:compSH}}
\bigskip
\begin{tabular}{@{\extracolsep{10pt}}*{7}{c}}
\hline\hline
$T_\text{e}$ & $N_\text{e}\text{(cm$^{-3}$)}$ & DLA & MOST & 
hybrid & detail & CA \\
\hline
25 & $10^{16}$ & 5.247 & 5.87 & 5.227 & 5.530 & 5.712 \\
   & $10^{18}$ & 6.050 & 6.17 & 6.101 & 6.352 & 6.406 \\
   & $10^{20}$ & 6.158 & 6.23 & 6.291 & 6.183 & 6.303 \\
   & $10^{22}$ & 3.403 & 3.78 & 4.200 & 3.083 & 3.248 \\[1ex]
50 & $10^{16}$ & 7.625 & 7.45 & 7.618 & 7.604 & 7.607 \\
   & $10^{18}$ & 7.739 & 7.80 & 7.797 & 7.791 & 7.791 \\
   & $10^{20}$ & 7.901 & 7.90 & 7.917 & 7.905 & 7.905 \\
   & $10^{22}$ & 6.275 & 6.31 & 6.706 & 6.134 & 6.194 \\   
\hline\hline
\end{tabular}
\end{center}
\end{table}

%\begin{table}[bh]%[htbp]
\begin{sidewaystable}[htbp]%[htbp]
\begin{center}
\caption{Accuracy check of the collisional-radiative (CR) 
\textit{detailed-level} solution for neon at 
$T_\text{e}=10\text{ eV}$. 
The average charge $\left<\Zs\right>$ is calculated using three 
kinds of microreversible rates: collisional rates 
($\left<\Zs_\text{coll}\right>$), radiative rates including 
absorption and stimulated emission in a Planckian field 
($\left<\Zs_\text{rad}\right>$), all these rates plus 
autoionization and dielectronic recombination rates 
($\left<\Zs_\text{all}\right>$). $\left<\Zs_\text{SB}\right>$ is 
the average charge at thermal equilibrium as derived from 
Saha-Boltzmann and should be equal to the previous 
$\left<\Zs\right>$ values at infinite numerical accuracy. 
The additional test $\delta_{\text{rad}}$ (resp. 
$\delta_{\text{col}}$, $\delta_{\text{all}}$,) is the maximum 
difference on the ion-level populations between the ``radiative'', 
(resp ``collisional'', ``complete'') system and the Saha-Boltzmann 
solution.\label{tab:acthsbNe}}
\bigskip
\begin{tabular}{@{\extracolsep{6pt}}*{8}{c}}
\hline\hline
 $N_e\text{(cm$^{-3}$)}$ & $\left<\Zs_\text{rad}\right>$ & 
 $\left<\Zs_\text{coll}\right>$ & $\left<\Zs_\text{all}\right>$ & 
 $\left<\Zs_\text{SB}\right>$ & $\delta_\text{rad}$ & 
 $\delta_\text{col}$ & $\delta_\text{all}$ \\
\hline
 $10^{12}$ & 7.712640377 & 7.712640378 & 7.712640377 & 7.712640377 & 
 $2.0\times10^{-10}$ & $2.7\times10^{-9}$ & $1.5\times10^{-10}$ \\
 $10^{14}$ & 6.639625837 & 6.639625830 & 6.639625838 & 6.639625838 & 
 $1.9\times10^{-10}$ & $3.4\times10^{-9}$ & $4.6\times10^{-11}$ \\
 $10^{16}$ & 5.491015460 & 5.491015458 & 5.491015465 & 5.491015460 & 
 $1.9\times10^{-12}$ & $7.2\times10^{-10}$ & $1.3\times10^{-9}$ \\
 $10^{18}$ & 4.260724652 & 4.260724657 & 4.260724656 & 4.260724652 & 
 $6.5\times10^{-13}$ & $1.5\times10^{-9}$ & $1.3\times10^{-9}$ \\
 $10^{20}$ & 2.847355968 & 2.847355980 & 2.847355979 & 2.847355968 & 
 $2.1\times10^{-14}$ & $1.1\times10^{-9}$ & $1.2\times10^{-9}$ \\
 $10^{22}$ & 0.719532736 & 0.719535868 & 0.719535378 & 0.719532736 & 
 $4.1\times10^{-14}$ & $6.9\times10^{-7}$ & $7.0\times10^{-7}$ \\
\hline\hline
\end{tabular}
\end{center}
\end{sidewaystable}
%\end{table}

%\enlargethispage{2.0cm}
\begin{table}[bh]%[htbp]
\begin{center}
\caption{Average ionization stage in a 10 eV-carbon plasma as a 
function of the electron density $N_\text{e}$. The last two 
columns are the configuration-average (CA) charge and the 
detailed-level charge derived from the collisional-radiative (CR) 
equations. The discrepancy between these values is correlated to 
the discrepancy between the Saha-Boltzmann charge (column 5) and 
the average charge derived from various modified rate-equation 
systems: collisional rates (column 2), radiative rates including 
a fictive Planckian field (column 3), all these rates plus 
autoionization (column 4). Data in columns 2--5 are obtained in the 
CA framework.\label{tab:acthsbC}}
\bigskip
\begin{tabular}{@{\extracolsep{5pt}} c c c c c@{\extracolsep{20pt}} c @{\extracolsep{10pt}} c}
\hline\hline
$N_\text{e}$& \multicolumn{4}{c}{Modified-rate system} & 
\multicolumn{2}{c}{CR system}\\[-5pt]
% Ne (cm-3)&  -------------- modified rate system -------------- 
%  -- full col-rad system -- \\
$(\text{cm}^{-3})$& $\left<\Zs_\text{coll}\right>$ & 
 $\left<\Zs_\text{rad}\right>$ & $\left<\Zs_\text{all}\right>$ & 
 $\left<\Zs_\text{SB}\right>$ & $\left<\Zs_\text{CA}\right>$ & 
  $\left<\Zs_\text{det}\right>$ \\
\hline
 $10^{12}$ &  4.000003693 & 4.000003671 & 4.000003671 & 4.000004022 & 3.7363 & 3.7441 \\
 $10^{14}$ &  3.999998275 & 3.999998275 & 3.999998275 & 3.999998278 & 3.7563 & 3.7616 \\
 $10^{16}$ &  3.999823807 & 3.999823806 & 3.999823806 & 3.999823805 & 3.9273 & 3.9273 \\
 $10^{18}$ &  3.982626435 & 3.982615120 & 3.982627955 & 3.982613441 & 3.9679 & 3.9679 \\
 $10^{20}$ &  3.225872703 & 3.197765566 & 3.230477092 & 3.194905628 & 3.2282 & 3.1902 \\
 $10^{22}$ &  1.052601006 & 0.689418858 & 1.069768540 & 0.717721378 & 1.0698 & 0.7113 \\
\hline\hline
\end{tabular}
\end{center}
\end{table}

\begin{table}[htpb]
\caption{Average ionization degree in a neon plasma at $T_\text{e}= 
30\text{ eV}$ with respect to the electron density $N_\text{e}$. 
Column 2 involves detailed levels, while columns 3--7 contain charges 
which are computed in the configuration % "which" added - postprint
average scheme. Columns 2 and 3 are the collisional-radiative 
solution, while columns 4--6 are variants of ``microreversibility 
check'', based on a collisional, radiative, or complete set of 
rates respectively. The closer these values are with respect to the 
Saha-Boltzmann $\left<\Zs\right>$ value (column 7), the better is 
assumed to be the configuration average. 
Column 8 is the average energy dispersion of levels inside a 
configuration with a ponderation by the collisional-radiative 
configuration populations.
\label{tab:CAchkNeT30Z}}
\begin{tabular}{c@{\bsp}c c c@{\ssp}cc@{\bsp}c@{\ssp}c@{\zsp}}
\hline\hline
$N_\text{e}$ & \multicolumn{2}{c}{CR system} & 
\multicolumn{3}{c}{Modified rate system} & Saha- & 
$\left<\Delta E_\text{CR}\right>$\\[-5pt]
$(\text{cm}^{-3})$ & Detailed & CA & Collisional & Radiative & 
Complete & Boltzmann & (eV)\\
\hline
% Te =  30 eV - Average Z* - Ne - colradconf_chth_bat / colradniv_1m_chth - version 2007/07
% Ne (cm-3)   Z* CR det   Z* CR       Z* coll      Z* radr       Z* allr       Z* SB   DeltaE CR(eV)
 $10^{12}$ & 5.7919 &  5.6236 &  8.000009396 & 8.000009460 & 8.000009460 & 8.000010081 &  0.069 \\  
 $10^{14}$ & 5.7911 &  5.6540 &  7.999998415 & 7.999998416 & 7.999998416 & 7.999998422 &  0.164 \\  
 $10^{16}$ & 6.2120 &  6.3134 &  7.999832154 & 7.999832154 & 7.999832154 & 7.999832154 &  2.18  \\  
 $10^{18}$ & 6.8256 &  6.8456 &  7.983410902 & 7.983409743 & 7.983409658 & 7.983407736 &  1.13  \\  
 $10^{20}$ & 6.9131 &  6.9400 &  7.180274366 & 7.172115334 & 7.180822653 & 7.168028134 &  0.97  \\  
 $10^{22}$ & 3.6140 &  3.6603 &  4.471505865 & 4.090408345 & 4.436454002 & 4.047538516 &  5.92  \\  
\hline\hline
\end{tabular}
\end{table}

\begin{table}
\caption{Second central moment 
$\left<(\Zs-\left<\Zs\right>)^2\right>$ in a neon plasma at % ^3 changed to ^2 - postprint 
$T_\text{e}= 30\text{ eV}$ as a function of $N_\text{e}$. Refer to 
Table \ref{tab:CAchkNeT30Z} for details. 
The average energy dispersion $\left<\Delta E_\text{CR}\right>$ is 
not repeated. \label{tab:CAchkNeT30m2}}
\begin{tabular}{c@{\bsp}c@{\ssp}c@{\bsp}c@{\ssp}cc@{\bsp}c@{\zsp}c}
\hline\hline
$N_\text{e}$ & \multicolumn{2}{c}{CR system} & 
\multicolumn{3}{c}{Modified rate system} & Saha- \\[-8pt]
$(\text{cm}^{-3})$ & Detailed & CA & Collisional & Radiative & Complete & Boltzmann \\
\hline
% Second central moment M2 = < (Z*-<Z*>)^2 >                                                
%           - detail -   --------------------- configuration average ---------------------
% Ne (cm-3)    M2 CR      M2 CR        M2 coll       M2 radr       M2 allr        M2 SB    
  $10^{12}$ &  0.6128 &  0.5283 &   0.000009430 & 0.000009494 & 0.000009494 & 0.000010114 \\
  $10^{14}$ &  0.6137 &  0.5433 &   0.000001773 & 0.000001773 & 0.000001773 & 0.000001780 \\
  $10^{16}$ &  0.5043 &  0.4528 &   0.000167828 & 0.000167828 & 0.000167828 & 0.000167829 \\
  $10^{18}$ &  0.3288 &  0.3137 &   0.016394148 & 0.016396461 & 0.016396557 & 0.016400426 \\
  $10^{20}$ &  0.4700 &  0.4377 &   0.442653567 & 0.456411086 & 0.440578198 & 0.459907235 \\
  $10^{22}$ &  0.7707 &  0.7484 &   0.785410789 & 0.787926179 & 0.811753013 & 0.823178341 \\
\hline\hline
\end{tabular}
\end{table}

\begin{table}
\caption{Third central moment 
$\left<(\Zs-\left<\Zs\right>)^3\right>$ in a neon plasma at 
$T_\text{e}= 30\text{ eV}$ as a function of $N_\text{e}$. Refer to 
Table \ref{tab:CAchkNeT30Z} for details. \label{tab:CAchkNeT30m3}}
\begin{tabular}{c@{\bsp}c@{\ssp}c@{\bsp}c@{\ssp}cc@{\bsp}c@{\zsp}c}
\hline\hline
$N_\text{e}$ & \multicolumn{2}{c}{CR system} & 
\multicolumn{3}{c}{Modified rate system} & Saha- \\[-8pt]
$(\text{cm}^{-3})$& Detailed & CA & Collisional & Radiative & 
Complete & Boltzmann \\\hline
% Third central moment  M3 = < (Z*-<Z*>)^3 >                                                
%          --- detail ---  --------------------- configuration average ---------------------
% Ne (cm-3)     M3 CR     M3 CR        M3 coll        M3 radr       M3 allr         M3 SB    
  $10^{12}$ & \m0.0647 & \m0.2229 & \m0.000009396 & \m0.000009460 & \m0.000009460 & \m0.000010080 \\
  $10^{14}$ & \m0.0685 & \m0.2146 & $-$0.000001585 & $-$0.000001584 & $-$0.000001584 & $-$0.000001578 \\
  $10^{16}$ & $-$0.0500 & $-$0.0416 &  $-$0.000167786 & $-$0.000167786 & $-$0.000167787 & $-$0.000167787 \\
  $10^{18}$ & $-$0.0289 & $-$0.0212 &  $-$0.016009480 & $-$0.016014093 & $-$0.016014161 & $-$0.016021878 \\
  $10^{20}$ & $-$0.0844 & $-$0.0508 &  $-$0.097745732 & $-$0.118986694 & $-$0.093018605 & $-$0.119420248 \\
  $10^{22}$ & \m0.2376 & \m0.3694 &  $-$0.125419985 & $-$0.024429621 & $-$0.109647319 & $-$0.010892532 \\
\hline\hline
\end{tabular}
\end{table}

\clearpage
%%%%%%%%%%%%%%%%%%%%%%%%%%%%%%%%%%%%%%%%%%%%%%%%%%%%%%%%%%%%%%%%%%
%%%%%%%%%%                   Figures                   %%%%%%%%%%%
%%%%%%%%%%%%%%%%%%%%%%%%%%%%%%%%%%%%%%%%%%%%%%%%%%%%%%%%%%%%%%%%%%
\enlargethispage{2.5cm}
\section*{Figures}

\end{document}